\documentclass[aps,prl,twocolumn,amsfonts,amsmath,%
groupedaddress]{revtex4-1}

\usepackage{ifpdf}
\ifpdf
	\usepackage[pdftex]{graphicx}
\else
	\usepackage[dvipdfm]{graphicx}
\fi
\usepackage{hyperref}

\newcommand{\bra}[1]{\left\langle #1 \right\vert}
\newcommand{\ket}[1]{\left\vert #1 \right\rangle}
\newcommand{\dg}{^{\circ}}

\newcommand{\mr}{\mathrm}
\newcommand{\rs}[1]{_{\mathrm{#1}}}
\newcommand{\sq}{_{\mathrm{sq}}}
\newcommand{\disp}{_{\mathrm{disp}}}
\newcommand{\dc}{_{\mathrm{dc}}}

\begin{document}

\title{Optical continuous-variable qubit}

\author{Jonas S. Neergaard-Nielsen}
\author{Makoto Takeuchi}
\author{Kentaro Wakui}
\author{Hiroki Takahashi}
\author{Kazuhiro Hayasaka}
\author{Masahiro Takeoka}
\author{Masahide Sasaki}
\affiliation{National Institute of Information and Communications Technology (NICT)\\
4-2-1 Nukui-kitamachi, Koganei, Tokyo 184-8795, Japan}

\date{\today}

\begin{abstract}
  In a new branch of quantum computing, information is encoded into
  coherent states, the primary carriers of optical communication. To
  exploit it, quantum bits of these coherent states are needed, but it
  is notoriously hard to make superpositions of such
  continuous-variable states. We have
  realized the complete engineering and characterization of a qubit of
  two optical continuous-variable states. Using squeezed vacuum as a
  resource and a special photon subtraction technique, we could with
  high precision prepare an arbitrary superposition of squeezed vacuum
  and a squeezed single photon. This could lead the way to
  demonstrations of coherent state quantum computing.
\end{abstract}

\maketitle

Among the various physical implementation schemes of quantum
information processing (QIP), optical QIP in traveling light fields is
a significant contender \cite{O'Brien2009}. Used at the nodes of a
quantum optical network, it would deliver ultra-high capacity with
minimum power in optical communications \cite{Waseda2010} and highly
secure communications and distributed QIP
\cite{Kimble2008,Briegel1998,Bennett1993} surpassing any classical
counterpart. Unfortunately, photons do not readily interact with each
other, making it quite a task to implement quantum gate elements. One
currently feasible scheme is linear optical quantum computing (LOQC),
which uses off-line resource states, linear optical processing, and
photon-number resolving detection
\cite{Knill2001,Gottesman2001}.%
There are two approaches to LOQC, the standard one being the single
photon scheme, where single photons are used as the physical quantum
bits (qubits) \cite{Knill2001}. The other is referred to as coherent
state quantum computing (CSQC), where two phase-opposite coherent
states are used for the qubits, i.e. $\ket{\uparrow}=\ket{\alpha}$,
$\ket{\downarrow}=\ket{-\alpha}$
\cite{Cochrane1999,Jeong2002,Ralph2003,Jeong2007b,Lund2008}.

CSQC is not only effective for exponential speed-up of computations,
but also for attaining the ultimate capacity of an optical channel in
current network infrastructure where coherent states are the primary
carriers. Because coherent states propagate intact, even through lossy
channels, simple coherent state-encoding is found to be the optimal
transmission strategy \cite{Giovannetti2004}. At the same time, the
optimal decoding should be fully quantum, and can be implemented by an
extension of CSQC \cite{Sasaki1998}. Practical implementation of CSQC
is still a big challenge, though -- one requirement is the availability
of arbitrary qubit states as resources. So far, two diagonal states of
the qubit $\ket{\alpha} \pm \ket{-\alpha}$ -- so-called Schr\"odinger
cat states -- have been generated in the laboratories
\cite{Ourjoumtsev2006, Neergaard-Nielsen2006, Wakui2007,
Ourjoumtsev2007a, Takahashi2008}.

We have implemented a setup that is suited for the generation of such
arbitrary qubits. For this demonstration, though, we perform the
complete engineering of a different, but closely related kind of qubit,
namely one with squeezed vacuum and squeezed single-photon states as
the basis. The squeezed photon state is in fact very similar to one of
the CSQC diagonal qubits, as was utilized in the previous
demonstrations of those. To create the arbitrary superposition of the
basis states, we use single-photon subtraction from a squeezed vacuum
assisted by a coherent displacement operation \cite{Takeoka2007}.
Photon subtraction is a simple, but powerful technique that has been
used for non-classical state generation \cite{Ourjoumtsev2006,
Neergaard-Nielsen2006, Wakui2007, Ourjoumtsev2007a, Takahashi2008},
entanglement increase \cite{Ourjoumtsev2007,Takahashi2010}, and
fundamental tests of quantum mechanics \cite{Zavatta2009} (see
\cite{Kim2008a} for an overview). A related experiment for engineering
of superpositions of the 0-, 1-, and 2-photon states was recently
reported \cite{Bimbard2009}. In our scheme, in contrast, the state
elements of the superposition lie in the infinite dimensional Hilbert
space, including many-photon number states. We should also note that
spectacular progress in generation of complex, high-photon number
states has been made in microwave resonator fields
\cite{Deleglise2008,Hofheinz2009}, but these states are trapped and
therefore of limited usefulness.

\begin{figure}
\begin{center}
\includegraphics[width=\linewidth]{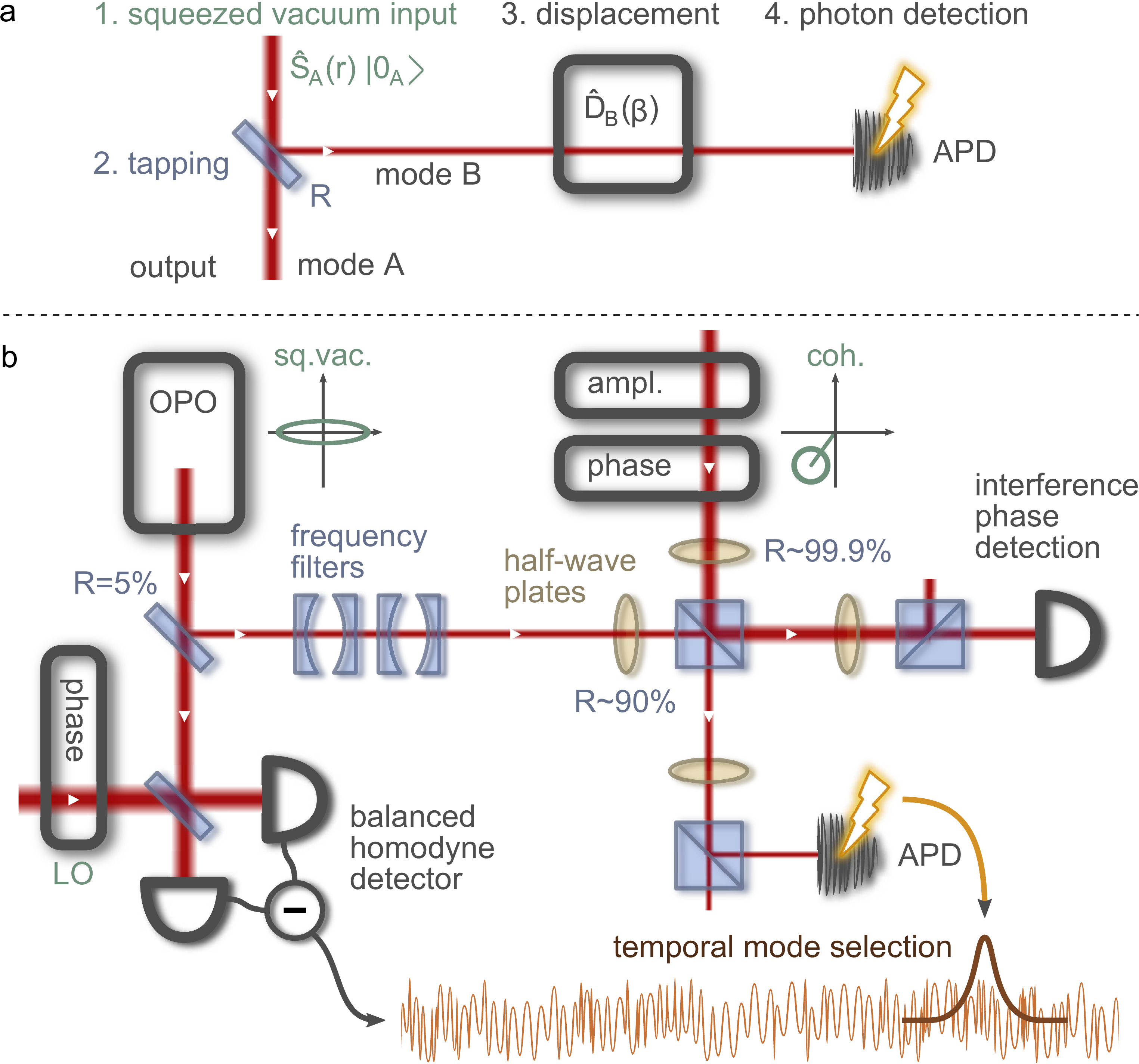}
\end{center}
\caption{\textbf{Experiment outline.}
\textbf{a)} Conceptual schematic. \textbf{b)} Experimental
setup. OPO, optical parametric oscillator; R, beam splitter
reflectivity; LO, local oscillator; APD, avalanche photo diode.}
\label{fig:Experimental_setup}
\end{figure}

A simplified schematic is shown in Fig. \ref{fig:Experimental_setup}a.
The squeezed vacuum state $\hat S_A(r)\ket{0_A}$ is prepared in mode A,
where $\hat S_A(r)$ represents the squeezing operation with strength
$r$. A small fraction $R$ of it is tapped off via a beam splitter as a
trigger beam in mode B, subjected to the displacement operation $\hat
D_B(\beta)$, and detected on an avalanche photodiode (APD). Without the
displacement operation, the output state conditioned on a click at the
APD would be
\begin{align}
  \bra{1_B} \hat V_{AB} \hat S_A(r) \ket{0_{AB}}
  \approx 
  -\sqrt{R} \sinh r \hat S_A(r) \ket{1_A},
\end{align}
where $\hat V_{AB}$ is the beam splitting operator. That is, a photon
has been subtracted from the squeezed vacuum state in mode A, which is
equivalent to squeezing a single photon. With the inclusion of the
displacement operation, this changes to
\begin{multline}\label{ideal displaced photon subtraction}
  \bra{1_B} \hat D_B(\beta) \hat V_{AB} \hat S_A(r) \ket{0_{AB}}
  \approx \\ 
  \mathcal{N}\big(\beta \hat S_A(r) \ket{0_A} -\sqrt{R} \sinh r \hat
  S_A(r) \ket{1_A} \big),
\end{multline}
with normalization factor $\mathcal{N}$. This superposition originates
from the two indistinguishable processes; an APD click comes either
from the displacement or from the squeezing, corresponding to the
output states $\hat S(r)\ket{0}$ or $\hat S(r)\ket{1}$, respectively. A
displacement photon is uncorrelated with the output mode, hence leaving
the squeezed vacuum state intact. The superposition weight and phase
can be completely controlled by the displacement operation. Each of the
two states is composed of several Fock state elements -- only even
photon numbers for $\hat S(r)\ket{0}$ and odd numbers for $\hat
S(r)\ket{1}$. They are orthogonal to each other, so together they
constitute a qubit basis and the general output state (\ref{ideal
displaced photon subtraction}) can therefore be represented on a Bloch
sphere as
\begin{equation}\label{ket_rho(theta,varphi)}
\ket{\rho(\theta,\varphi)} =
\cos\frac{\theta}{2} \hat S(r) \ket{0} + e^{i\varphi}
\sin\frac{\theta}{2} \hat S(r) \ket{1} ,
\end{equation}
with $\varphi = \pi - \arg\beta$ and, after normalization,
\begin{equation}\label{theta_clickrate_relation}
\cos\frac{\theta}{2} = \frac{|\beta|}{\sqrt{|\beta|^2 + R\sinh^2r}} =
\sqrt{\frac{n_{\mathrm{disp}}}{n_{\mathrm{disp}} + n_{\mathrm{sq}}}} ,
\end{equation}
where $n_{\mathrm{disp}}$ ($n_{\mathrm{sq}}$) are the number of photons
in mode $B$ originating from the displacement beam (squeezing).

Our experimental setup is shown in detail in Fig.
\ref{fig:Experimental_setup}b. The input squeezed vacuum states are in
a cw beam generated by an optical parametric oscillator (OPO) at a
center wavelength of 860nm with a bandwidth $\zeta_0/\pi\sim9$\,MHz
(FWHM). This cw beam is intuitively a continuous sequence of squeezed
light packets, each of which is in a temporal form $
\psi(t)=\sqrt{\zeta_0} e^{-\zeta_0 \vert t\vert} $ (the squeezing level
within these packets is -2.7 dB in our experiment).

After tapping off $R=5\%$ for the trigger, the main part of the light
(the output signal) is measured on a homodyne detector for state
analysis. The trigger beam is spectrally filtered by two subsequent
Fabry-Perot resonators before it is displaced and directed onto a Si
APD. The phase space displacement is implemented by overlapping the
beam with a weak coherent state on an imbalanced beam splitter
\cite{Paris1996}. For experimental convenience, we use a combination of
half-wave plates (HWP) and a polarizing beam splitter (PBS) which
allows for independent tuning of the splitting ratios of the two beams.
For the trigger beam, 90\% is reflected towards the APD, while only
$\sim$0.1\% of the displacement beam is transmitted in that direction.
The other output mode of the PBS is monitored on a standard linear
photodiode for the purpose of locking the displacement phase. The phase
monitoring is done with the help of chopped probe light (10 kHz, 20\%
duty cycle) in both the OPO output and the displacement beam modes, and
the locking is controlled by FPGA modules. Apart from this essential
displacement part, the setup is mostly identical to that in Refs.
\cite{Wakui2007,Takahashi2008}, where more details are provided.

The time resolution of the trigger signal is on a scale of sub ns,
which is almost instantaneous compared with the time scale of the
squeezed packets ($\pi/\zeta_0\sim100$\,ns). The trigger signal, say at
time $t_1$, specifies a temporal mode $\psi(t-t_1)$ of the output state
of interest. In the homodyne channel, a continuous photo-current is
sampled around the trigger time $t_1$, and subsequently integrated over
this packet mode function to yield the observed quadrature variable of
the field. The specific quadrature to be measured is determined by the
phase of the local oscillator (LO) of the homodyne detector. To obtain
full information about the output quantum state, we must carry out many
homodyne measurements at a range of different LO phases, that is, a
tomographic state reconstruction. For each state, 360,000 quadrature
points were observed, distributed on 12 different LO phases. The
reconstruction was then done by the maximum likelihood method
\cite{Lvovsky2009} without any correction of measurement losses.

\begin{figure*}[htb]
\begin{center}
\includegraphics[width=\textwidth]{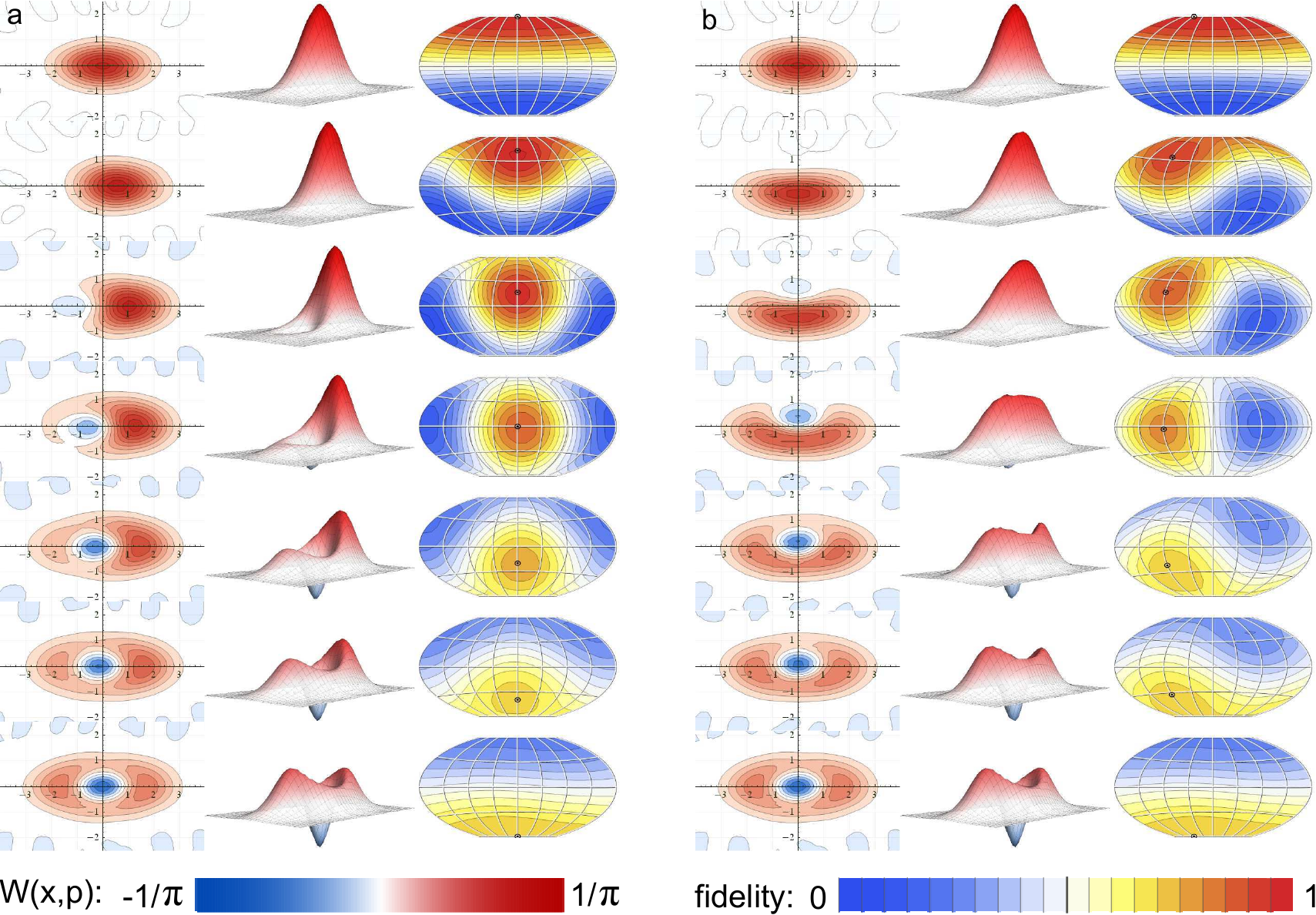}
\end{center}
\caption{\textbf{Control of superposition weight.} A variety of
experimentally generated superposition states where the superposition
weight $\theta$ has been swept with the phase $\phi$ fixed at
\textbf{a)} 0$^{\circ}$, and \textbf{b)} $-90^{\circ}$. The left and
center panels show the Wigner function as a contour plot and surface
plot, respectively. The axes are the $x$- and $p$-quadratures. The
right panels show a flattened Bloch sphere with an overlay signifying
the fidelity between an ideal qubit (eq. \ref{ket_rho(theta,varphi)})
at the given $(\theta,\varphi)$ point and the measured state. The
central longitude corresponds to $\varphi=0\dg$, and $\theta=0\dg$ at
the north pole. The small circles serve to point out which state we
were aiming at. The ideal qubit is taken to have a squeezing parameter
$r = 0.38$.} \label{fig:Longitude_sweep}
\end{figure*}

\begin{figure*}[htb]
\begin{center}
\includegraphics[width=\textwidth]{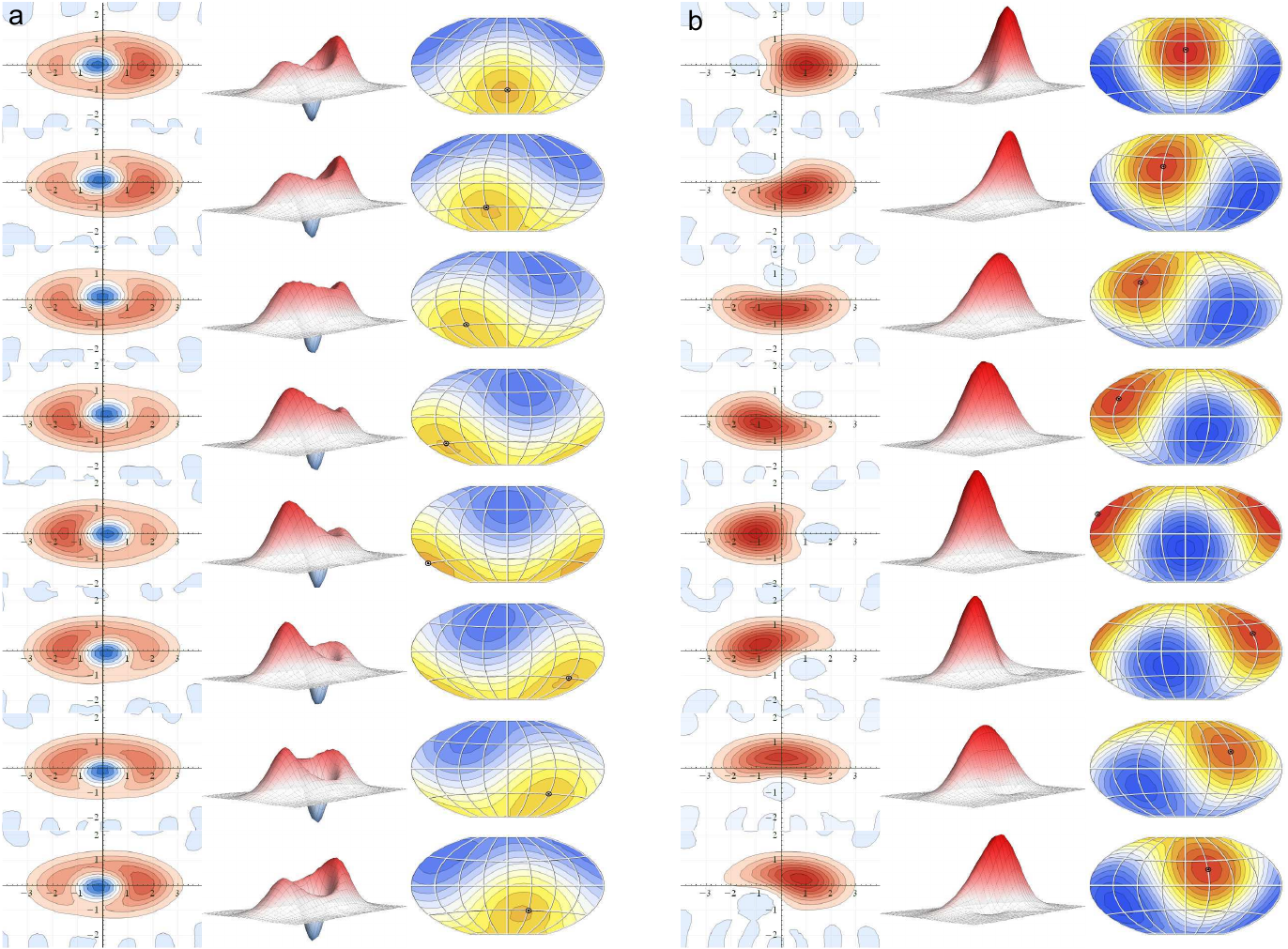}
\end{center}
\caption{\textbf{Control of superposition phase.} Generated states with
the superposition phase $\phi$ swept while keeping the superposition
weight $\theta$ constant at \textbf{a)} 135$^{\circ}$, and \textbf{b)}
60$^{\circ}$.} \label{fig:Phase_sweep}
\end{figure*}

A selection of our generated states are presented in Figs.
\ref{fig:Longitude_sweep} and \ref{fig:Phase_sweep}. Each state is
represented by its Wigner function as a top-down contour plot and a 3D
view, as well as by a Bloch sphere map of the fidelity between the
state and the ideal squeezed qubit (\ref{ket_rho(theta,varphi)}) for
all combinations of $\theta$ and $\phi$. To demonstrate the performance
of the state engineering, we show in Fig. \ref{fig:Longitude_sweep} the
control of the superposition weight $\theta$ while keeping the phase
constant at a) $\varphi=0\dg$ and b) $\varphi=-90\dg$, and conversely,
in Fig. \ref{fig:Phase_sweep} we show the control of the complex phase
$\varphi$ for fixed weights of the superposition, a) $\theta =
135^{\circ}$ and b) $\theta = 60^{\circ}$. In Fig.
\ref{fig:Longitude_sweep} we see that by increasing the amount of
displacement in the trigger channel, $|\beta|$, we can move from the
south pole (squeezed photon) to the north pole (squeezed vacuum) of the
Bloch sphere along a fixed longitude. While doing that, the negative
dip of the Wigner function moves away from the center in a direction
determined by the displacement phase. At the same time, the dip becomes
shallower and finally disappears as the state approaches the north
pole. In Fig. \ref{fig:Phase_sweep}, on the other hand, when sweeping
the displacement phase, $\mathrm{arg}\beta$, the negative dip circles
around the center while the state on the Bloch sphere turns around at a
fixed latitude.

From the fidelity maps, we can see that there is a clearly defined
qubit state of maximum fidelity in the center of the orange parts of
the maps. These maximum points are all quite close to the target states
that we aimed for in each qubit realization -- these targets are marked
by small circles on the fidelity maps. This illustrates the precision
of our state control. It is also clear that the obtained fidelities are
not as high around the south pole as they are in the north. That is
because highly non-classical states with negative Wigner functions,
such as the squeezed single photon, are much more fragile and
susceptible to losses than, for example, squeezed vacuum.

\begin{figure}
\begin{center}
\includegraphics[width=\linewidth]{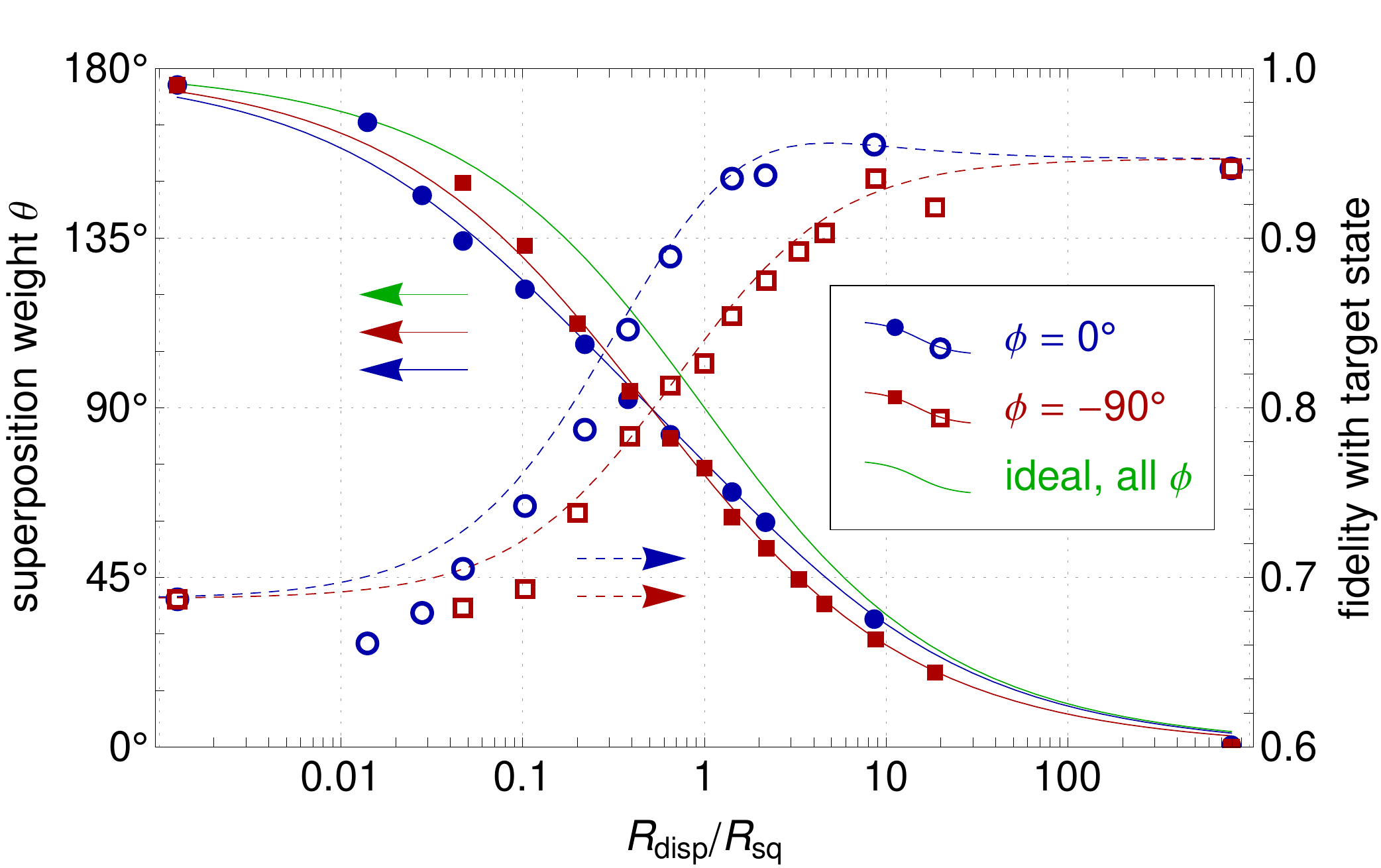}
\end{center}
\caption{\textbf{Influence of displacement strength.}
Experimentally obtained superposition weights (filled points + solid
curves, left axis) and fidelities with the intended target states
(unfilled points + dashed curves, right axis) versus the APD count rate
of displacement photons relative to squeezing photons for two series of
measurements with different superposition phase factor. The curves are
obtained from a theoretical model with no free parameters, taking into
account the various imperfections. The green curve is the relation for
the idealized, loss-less case. Note that the extreme data points
correspond to count rate ratios of 0 and infinity.}
\label{fig:click_ratio}
\end{figure}

Equation (\ref{theta_clickrate_relation}) anticipates a direct
correspondence between the APD click rates and the resulting
superposition weight. The photon number from squeezing (displacement)
is directly proportional to the count rate $R_{\mathrm{sq}}$
($R_{\mathrm{disp}}$) observed when blocking the displacement (trigger)
beam before the overlapping PBS, so the relation would read $\theta =
2\tan^{-1}(R_{\mathrm{disp}}/R_{\mathrm{sq}})^{-1/2}$. This relation is
shown as the green curve in Fig. \ref{fig:click_ratio}, where also the
experimentally obtained $\theta$ (of the ideal qubit with maximum
fidelity) as a function of the click rate ratios
$R_{\mathrm{disp}}/R_{\mathrm{sq}}$ are plotted. The data points are
shown for both the $\varphi=0\dg$ and the $\varphi=-90\dg$ series of
state generation. All the points are lying below the theoretical curve.
That is because the model in eqs. (\ref{ideal displaced photon
subtraction}--\ref{theta_clickrate_relation}) is an idealized picture
from which there are several deviations in the actual experiment. In
particular, the unavoidable losses in state generation and measurement
serve to mix in vacuum which effectively acts to decrease the obtained
$\theta$. To more accurately describe the experimental outcomes, we
have developed a relatively simple model (see the appendix) which takes
into account a more detailed description of the OPO and the trigger
filtering, as well as all the inefficiencies of the setup. This model,
with no free parameters (except for the squeezing parameter $r$ that is
semi-fixed), is also plotted in Fig. \ref{fig:click_ratio} and is seen
to simulate the measured outcomes very well. Apart from the
superposition weights, the figure also shows the fidelities between the
measured states and the target states -- that is, the fidelity values
at the target marks in the Bloch sphere maps of Fig.
\ref{fig:Longitude_sweep}. The state preparation works somewhat better
for displacement along the anti-squeezing direction ($\phi=0°$) than
along the squeezing direction. This can be ascribed to the fact that
losses have a larger influence on the squeezed than on the
anti-squeezed quadrature. The fidelities are also well described by the
theoretical model, although with some smaller discrepancies in the
weak-displacement regime.

As these results show, we were able to realize with high precision and
relatively high fidelity the complete engineering of a qubit of
continuous-variable states. The simple superposition preparation
technique demonstrated here can be straight-forwardly applied to
generation of coherent state qubits. If the input state instead of
squeezed vacuum were one of the `cat' states
$\ket{\alpha}\pm\ket{-\alpha}$, then the single photon subtraction
would change this input into the opposite state
$\ket{\alpha}\mp\ket{-\alpha}$. Thus, combining this technique with the
already existing cat state \cite{Takahashi2008}, we should be able to
generate an arbitrary superposition of these two cat states. This is
equivalent to arbitrary superpositions of the coherent states
$\ket{\alpha}$ and $\ket{-\alpha}$, i.e. the coherent state qubits that
are cornerstones of CSQC. As a side-note, the currently measured basis
states, squeezed vacuum and squeezed photon, already have a quite good
resemblance with the `+'-cat and `-'-cat states with fidelities of 81\%
and 68\%, respectively, for a cat-amplitude $\alpha=1.0$. It is also
possible to generate larger amplitude cat states by cascading the
photon-subtraction \cite{Nielsen2007a}.
The results reported here can therefore lead the way to prototypes of
CSQC quantum gates, likely to become important ingredients for
attaining the ultimate capacity in future quantum optical networks.

\textbf{Acknowledgments} We would like to acknowledge helpful
discussions with Hyunseok Jeong and Chang-Woo Lee.

\bibliography{tiger-nourl}

\begin{thebibliography}{1}
\bibitem{app:Note1}
  \bibinfo {note} {The trigger filtering consists of two Fabry-Perot cavities,
  each with a Lorentzian transmission spectrum. However, one cavity is
  considerably narrower than the other, so we model the total effect of the
  filtering as a single Lorentzian filter with a bandwidth slightly lower than
  that of the narrowest cavity.}%
\bibitem{app:Note2}
  \bibinfo {note} {The experimental reconstruction method actually gives a
  density matrix instead of a Wigner function, so for the measured states we
  calculate fidelity by the trace of the density matrix products instead. This
  is equivalent to the Wigner function overlap.}%
\end{thebibliography}

\clearpage
\appendix

\begin{center}
\textbf{\large Appendix: Theoretical model}
\end{center}

In this appendix, we outline the theoretical modeling of the quantum
states generated in our experiment -- a model that fits very well with
the observations, as seen in Fig. 4 of the main paper.

We start by recounting the experimental setup and laying out all the
relevant parameters. Next, we describe a general Wigner function
formalism for calculating the conditional states and relate it to our
setting, including imperfections of the displaced photon subtraction.
Finally, we specify the actual input Gaussian state, determined by the
OPO output correlations, the trigger channel filtering, and our choice
of temporal mode functions. To round off, we also include the Wigner
function for the ideal squeezed vacuum/squeezed photon qubit -- the
class of states that we match fidelities against.

Other relevant and more general theoretical descriptions can be found
in Refs. \cite{app:Suzuki2006,app:Sasaki2006a,app:Takeoka2007}.

\begin{table*}
  \centering
\begin{tabular}{lll}
  \hline
  $\gamma$    & OPO bandwidth    & 2$\pi\times$4.5\rm{ MHz} \\
  $\epsilon$  & pump level       & variable -- in this paper, 0.3$\gamma$ \\
  $\kappa$    & filter bandwidth & $\sim 2\pi\times$25\rm{ MHz} \\
  \hline
  $T_t$       & tapping BS transmission     & 0.95 \\
  $\eta_A$    & overall efficiency, signal  & 0.82 ($0.96\times 0.91\times 0.98^2\times 0.99\times 0.99$) \\
  $\eta_B$    & overall efficiency, trigger & $\sim$ 0.1 \\
  \hline
  $R\sq$      & click rate, squeezed photons     & variable -- in this paper, 3600\rm{ c/s} \\
  $R\dc$      & click rate, displacement photons & $\sim$30\rm{ c/s} \\
  $R\disp$    & click rate, dark counts          & variable -- typically 50-50,000\rm{ c/s} \\
  \hline
  $\phi\disp$ & displacement angle         & variable \\
  $\chi$      & displacement mode matching & 0.97 \\
  \hline
\end{tabular}
  \caption{Model parameters and typical experimental values.
  The factors of the signal efficiency are: OPO escape, propagation, LO
  visibility, quantum efficiency, dark noise.}
  \label{tbl:modelparameters}
\end{table*}

\section{EXPERIMENTAL SETTING}

Initially, a cw squeezed vacuum is generated by a sub-threshold OPO
into mode A. The OPO HWHM bandwidth is $\gamma$ and the pump level is
$\epsilon<\gamma$. We align the phase space to have squeezing in the
$p$-quadrature. The normal ordered temporal output correlations are
then \cite{app:Gardiner2000}
\begin{gather}\label{eq:OPOcorrelations}
  \langle :\! \Delta \hat x\rs{A,ini}(t) \Delta \hat x\rs{A,ini}(t') \!: \rangle =
  \frac{\gamma\epsilon}{\gamma-\epsilon} e^{-(\gamma-\epsilon)|t-t'|}
  ,\\ \label{eq:OPOcorrelations2}
  \langle :\! \Delta \hat p\rs{A,ini}(t) \Delta \hat p\rs{A,ini}(t') \!: \rangle =
  -\frac{\gamma\epsilon}{\gamma+\epsilon} e^{-(\gamma+\epsilon)|t-t'|} .
\end{gather}
The squeezed vacuum is split on a beam splitter with transmission
$T_t$, being mixed with vacuum from mode B. The transmitted beam is the
output signal to be recorded by the homodyne detector whose overall
efficiency we denote by $\eta_A$ -- this efficiency parameter includes
internal OPO losses, propagation losses etc. The reflection is sent
towards an APD, being displaced on the way (after frequency filtering)
by mixing it with a coherent beam. The details of the displacement
process are not important: What matters is the rate of APD clicks
originating from the squeezed light, $R\sq$, or from the displacement
beam, $R\disp$ -- a click coming from the squeezed light heralds a
photon subtraction in mode A, while a click coming from the
displacement beam (uncorrelated with mode A) heralds no action.

Other important parameters are the spatial mode matching between
trigger and displacement beam, $\chi$, the dark count rate, $R\dc$, the
overall efficiency of mode B (trigger), $\eta_B$, the trigger filter
bandwidth \cite{app:Note1}, $\kappa$, and the phase angle of
displacement, $\phi\disp$. The spatial mode matching $\chi$ must be
close to 1. If not, the two possible origins of the detected photon are
not indistinguishable, leading to a mixed state instead of a
superposition.

The parameters used in this model together with their typical
experimental values are listed in Table \ref{tbl:modelparameters}.
Apart from these, there are also the parameters that define the signal
mode function as used in the extraction of the homodyne data:
$\gamma_f$, $\epsilon_f$, and $\kappa_f$. These can be chosen freely,
but we usually use the experimentally-equivalent values, that is,
$\gamma_f=\gamma$, $\epsilon_f=\epsilon$, $\kappa_f=\kappa$.

\section{GENERAL OUTPUT STATE FORMALISM}

To describe the effect of the photon subtraction, we adopt the Wigner
formalism of Ref. \cite{app:Molmer2006}. The two-mode state prior to
the photon detection event is Gaussian, i.e. it has a Gaussian Wigner
function fully determined by the covariance matrix $\Gamma$ and
displacement vector $\mathbf{d}$:
\begin{equation}\label{eq:Gaussian_Wigner}
  W_G(x_A,p_A,x_B,p_B) = \frac{1}{\pi^2\sqrt{\det\Gamma}}
    e^{-(\mathbf{x}-\mathbf{d})^T \Gamma^{-1} (\mathbf{x}-\mathbf{d})} ,
\end{equation}
with $\mathbf{x} = (x_A,p_A,x_B,p_B)^T$.

To get the state conditional on an APD click in mode B, the Gaussian
Wigner function should be multiplied by the Wigner function
corresponding to the detection operation and integrated over mode B --
this corresponds to $\mr{tr}_B[\hat{\rho}_{AB}\hat{\Pi}_B]$ in the
density matrix formulation. For the non-photon number resolving APD, we
use the standard on/off operation $\hat\Pi\rs{on} =
\hat{I}-\ket{0}\bra{0}$, which has the equivalent Wigner function
\begin{equation}
  W\rs{on}(x,p) = W\rs{id}(x,p) - W\rs{vac}(x,p) = \frac{1}{2\pi} -
    \frac{1}{\pi}e^{-x^2-p^2} .
\end{equation}
The output state conditioned on a displaced 1-photon subtraction is
therefore
\begin{multline}
  W\rs{d-1ps}(x_A,p_A) = \\
  \mathcal{N}\rs{out}\, 2\pi \!\! \iint_{-\infty}^{\infty}
   \hspace{-10pt} W_G(x_A,p_A,x_B,p_B) W\rs{on}(x_B,p_B)\, dx_Bdp_B ,
\end{multline}
with normalization to be determined by integrating the output state.

\subsection{Including dark counts and mode matching}

If an APD click is a dark count or if it comes from the part of the
displacement beam that is not mode-matched to the trigger beam, there
is no action on the signal beam and the output state is simply a
squeezed vacuum state
\begin{multline}
  W\rs{sq}(x_A,p_A) = \\ 2\pi \!\! \iint_{-\infty}^{\infty}
   \hspace{-10pt}  W_G(x_A,p_A,x_B,p_B) W\rs{id}(x_B,p_B) \, dx_Bdp_B \\
  =  \iint_{-\infty}^{\infty} \hspace{-10pt} W_G(x_A,p_A,x_B,p_B) \, dx_Bdp_B .
\end{multline}
If, on the other hand, the click originates from the trigger beam, but
from the part which is not mode-matched to the displacement beam, the
result will be a normal un-displaced photon subtraction:
\begin{multline}
  W\rs{1ps}(x_A,p_A) = \\
  \mathcal{N}\rs{out}\, 2\pi \!\! \iint_{-\infty}^{\infty}
   \hspace{-10pt} W\rs{G,nd}(x_A,p_A,x_B,p_B) W\rs{on}(x_B,p_B)\,
   dx_Bdp_B .
\end{multline}
with the un-displaced Gaussian state
\begin{equation}
  W\rs{G,nd}(x_A,p_A,x_B,p_B) = \frac{1}{\pi^2\sqrt{\det\Gamma}}
    e^{-\mathbf{x}^T \Gamma^{-1} \mathbf{x}} \ .
\end{equation}

Let the total count rate be $R=R\sq+R\disp+R\dc$. The final output
state is then (with $(x_A,p_A)$ temporarily removed)
\begin{widetext}
\begin{equation}\label{eq:Wout}
\begin{split}
  W\rs{out} &= \frac{R\sq+R\disp}{R} \bigg(
    \chi W\rs{d-1ps} + (1-\chi) \bigg(
    \frac{R\disp}{R\sq+R\disp} W\rs{sq} \ +
    \frac{R\sq}{R\sq+R\disp} W\rs{1ps}
    \bigg) \bigg)  +  \frac{R\dc}{R} W\rs{sq} \\
    &= \chi \frac{R\sq+R\disp}{R} W\rs{d-1ps} +
    (1-\chi) \frac{R\sq}{R} W\rs{1ps} +
    \bigg( (1-\chi) \frac{R\disp}{R} + \frac{R\dc}{R} \bigg)
    W\rs{sq} \ .
\end{split}
\end{equation}
\end{widetext}

\subsection{Explicit Wigner function}

Any two-mode Gaussian state can -- via local operations -- be put on
the following generic form:
\begin{equation}
  \Gamma = \left( \begin{array}{cccc}
                    a & 0 & e & 0 \\
                    0 & b & 0 & f \\
                    e & 0 & c & 0 \\
                    0 & f & 0 & d \\
                  \end{array} \right) ,\quad
  \mathbf{d} = \left( \begin{array}{c}
                        r \\ s \\ t \\ u \\
                      \end{array} \right) .
\end{equation}
By having aligned the squeezing quadrature of the initial state along
one of the phase space axes, we have already obtained this form (see
next section), where the $x$ and $p$ variables are completely
decoupled. We can set $r=s=0$ as displacement only occurs in mode B.

When inserting $\Gamma$ and $\mathbf{d}$ in \eqref{eq:Gaussian_Wigner}
and carrying out the integrations, we get the following normalized
expressions for the three state components of \eqref{eq:Wout}:
\begin{align}
  W\rs{sq}(x_A,p_A) &= \frac{1}{\pi\sqrt{ab}}
    e^{ -\frac{1}{a} x_A^2 -\frac{1}{b} p_A^2} , \\
  W\rs{1ps}(x_A,p_A) &= \frac{1}{1-w} \frac{1}{\pi\sqrt{ab}} e^{
    -\frac{1}{a} x_A^2 - \frac{1}{b} p_A^2}  \nonumber \\
    &\quad - \frac{w}{1-w}
    \frac{1}{\pi\sqrt{a'b'}} e^{ -\frac{1}{a'} x_A^2 -
    \frac{1}{b'} p_A^2 } , \\
  W\rs{d-1ps}(x_A,p_A) &= \frac{1}{1-w_d} \frac{1}{\pi\sqrt{ab}} e^{
    -\frac{1}{a} x_A^2 - \frac{1}{b} p_A^2}  \nonumber \\
    &\hspace{-43pt} - \frac{w_d}{1-w_d}
    \frac{1}{\pi\sqrt{a'b'}} e^{ -\frac{1}{a'}(x_A-r_d)^2 -
    \frac{1}{b'}(p_A-s_d)^2 } ,
\end{align}
with
\begin{eqnarray}
  a' &=& a-\frac{e^2}{1+c} ,\quad b' = b-\frac{f^2}{1+d} , \\
  r_d &=& -\frac{et}{1+c} ,\quad s_d = -\frac{fu}{1+d} ,
\end{eqnarray}
\begin{eqnarray}
  w &=& \frac{2}{\sqrt{(1+c)(1+d)}} , \\
    w_d &=& \frac{2}{\sqrt{(1+c)(1+d)}} e^{-t^2/(1+c) - u^2/(1+d)} .
\end{eqnarray}

The photon subtracted states are just the difference between two
squeezed states. For the total state, we can then gather terms to get
\begin{widetext}
\begin{equation}\label{eq:Wout2}
\begin{split}
  W\rs{out}(x_A,p_A) = &
    \bigg( \chi\frac{R\sq+R\disp}{R}\frac{1}{1-w_d}
          + (1-\chi)\frac{R\sq}{R}\frac{1}{1-w}
          + \frac{(1-\chi)R\disp+R\dc}{R} \bigg)
    \frac{1}{\pi\sqrt{ab}} e^{ -\frac{1}{a} x_A^2 -\frac{1}{b} p_A^2} \\
  & - (1-\chi)\frac{R\sq}{R}\frac{w}{1-w}
    \frac{1}{\pi\sqrt{a'b'}} e^{-\frac{1}{a'}x_A^2 -\frac{1}{b'}p_A^2}
  \ \ -\ \ \chi\frac{R\sq+R\disp}{R}\frac{w_d}{1-w_d}
    \frac{1}{\pi\sqrt{a'b'}} e^{ -\frac{1}{a'}(x_A-r_d)^2
                                 - \frac{1}{b'}(p_A-s_d)^2 } .
\end{split}
\end{equation}
\end{widetext}

\section{PHYSICAL DESCRIPTION OF STATE BEFORE TRIGGER DETECTION}

In this section, the generalized variables of the covariance matrix and
displacement vector in the previous section are related to experimental
parameters. The two-mode Gaussian state before the trigger detection
has temporal correlations given by the OPO output correlations
\eqref{eq:OPOcorrelations}-\eqref{eq:OPOcorrelations2}, by the trigger
filtering and by losses. The APD click automatically defines a
temporally localized single mode of the trigger beam, and based on the
click time, we define a localized mode for the signal beam. Although
this mode selection occurs after the actual photon subtraction, it is
possible in the model to impose it on the initial Gaussian state, in
order to get a time-independent covariance matrix and displacement
vector.

\subsection{Covariance matrix}

The initial time-dependent covariance matrix is modified by the tapping
beam splitter, losses, and the mode selection. The handling of
covariance matrix transformations is simplified a little by using
normal ordering - we do not have to take account of the vacuum
contributions until the end.

The initial two-mode time-dependent covariance matrix is
\begin{equation}
  :\! \Gamma\rs{OPO}(t-t') \!:
   = \left(
       \begin{array}{cccc}
         :\! \Gamma\rs{OPO,11} \!: & 0 & 0 & 0\\
         0 & :\! \Gamma\rs{OPO,22} \!: & 0 & 0 \\
         0 & 0 & 0 & 0 \\
         0 & 0 & 0 & 0 \\
       \end{array}
     \right) ,
\end{equation}
with
\begin{eqnarray}
  :\! \Gamma\rs{OPO,11} \!: &=& 2 \langle :\!
    \Delta \hat x\rs{A,ini}(t) \Delta \hat x\rs{A,ini}(t') \!: \rangle ,\\
  :\! \Gamma\rs{OPO,22} \!: &=& 2 \langle :\!
    \Delta \hat p\rs{A,ini}(t) \Delta \hat p\rs{A,ini}(t') \!: \rangle .
\end{eqnarray}
The tapping beam splitter, represented by the matrix $V(T_t)$,
transforms it to
\begin{equation}
  :\! \Gamma\rs{bs}(t-t') \!: =
    V(T_t) :\!\Gamma\rs{ini}(t-t')\!: V(T_t)^T .
\end{equation}

To obtain the state within the selected temporal mode, we integrate the
quadrature variables over the relevant filter functions, $f_A(t),
f_B(t)$, for example,
\begin{equation}
  \hat x_A = \int_{-\infty}^{\infty} f_A(t) \hat x\rs{A,bs}(t) dt .
\end{equation}
When calculating the elements of the covariance matrix $\Gamma$, we can
move the integration outside the expectation values, e.g.
\begin{equation}
\begin{split}
  :\! \Gamma_{13} \!: &= 2 \langle :\! \Delta \hat x_A \Delta \hat x_B \!: \rangle \\
    &= 2 \langle :\! \int\!\! f_A(t) \Delta \hat x\rs{A,bs}(t) dt
         \int\!\! f_B(t') \Delta \hat x\rs{B,bs}(t') dt' \!: \rangle \\
    &= \int\!\! f_A(t) f_B(t') \ 2 \langle :\! \Delta \hat x\rs{A,bs}(t)
         \Delta \hat x\rs{B,bs}(t') \!: \rangle dt dt' \\
    &= \int\!\! f_A(t) f_B(t') :\! \Gamma\rs{bs,13}(t-t') \!: dtdt' .
\end{split}
\end{equation}

Losses in both modes, as well as the trigger frequency filtering can be
incorporated in the filter functions (see later). Therefore, the final
matrix is
\begin{equation}
  :\! \Gamma \!:\ = \iint \mathbf{f}(t)^T :\! \Gamma\rs{bs}(t-t') \!:
    \mathbf{f}(t') \,dtdt'  ,
\end{equation}
with $\mathbf{f}(t) = \mathrm{diag}(f_A(t), f_A(t), f_B(t), f_B(t) )$,
and, remembering to put back the vacuum:
\begin{equation}
  \Gamma =\ :\! \Gamma \!: + \Gamma\rs{vac} .
\end{equation}

\subsection{Displacement vector}

With a displacement $\hat D_B(\beta) = \hat
D_B\big(|\beta|e^{i\phi\disp}\big)$ in mode B, the displacement vector
is
\begin{equation}
  \mathbf{d}=\sqrt{2}|\beta|\ (0,0,\cos\phi\disp,\sin\phi\disp)^T  .
\end{equation}
The photon number in the mode due to this displacement is of course
$n\disp=|\beta|^2$. On the other hand, from $n=(\langle :\! \Delta \hat
x^2 \!: \rangle + \langle :\! \Delta \hat p^2 \!: \rangle)/2$, the
photon number due to the squeezing can be obtained as
\begin{equation}
  n\sq = \frac{1}{4} \left( :\! \Gamma_{33} \!: + :\! \Gamma_{44} \!: \right) .
\end{equation}

The photon numbers are not directly experimentally accessible, but they
are proportional to the detected APD click rates, so we can get
$|\beta|$ and the displacement vector in terms of these rates:
\begin{equation}
  |\beta| = \sqrt{\frac{n\disp}{n\sq} n\sq}
    = \sqrt{ \frac{R\disp}{R\sq}
         \frac{:\! \Gamma_{33} \!: + :\! \Gamma_{44} \!:}{4} } \ ,
\end{equation}
\begin{equation}
  \mathbf{d} = \sqrt{ \frac{R\disp}{R\sq} \frac{:\! \Gamma_{33} \!: +
    :\! \Gamma_{44} \!:}{2} }\ (0,0,\cos\phi\disp,\sin\phi\disp)^T \ .
\end{equation}

\subsection{Mode functions}

In the approach taken here, the filter functions play two different
roles:
\begin{enumerate}
  \item To model the physical effect on the correlation matrix by
      losses and trigger filtering.
  \item To describe the actual observed temporal mode.
\end{enumerate}
\#2 is the original meaning of the mode function, but \#1 can be
included very nicely.

For the signal (mode A), the observed mode is the temporal filter
applied to the raw homodyne data in post-processing. This can be chosen
freely, but we always use the experimentally most reasonable function
which is
\begin{equation}
  \psi_A(t) = \sqrt{\mathcal{N}_A} \left( \frac{e^{-\gamma|t|}}{\gamma} -
    \frac{e^{-\kappa|t|}}{\kappa} \right) ,
\end{equation}
with normalization $\mathcal{N}_A = \gamma^3\kappa^3(\gamma+\kappa) /
(\gamma^4 + \gamma^3\kappa - 4\gamma^2\kappa^2 + \gamma\kappa^3 +
\kappa^4)$. This is the double-sided exponential from the OPO output
correlations, smoothened by the frequency filtering of the trigger.
Including the overall signal efficiency gives the filter function
\begin{equation}
  f_A(t) = \sqrt{\eta_A}\psi_A(t) .
\end{equation}

The APD detection time is very short relative to the correlation time,
so the temporal mode can be taken to be a delta function:
\begin{equation}
  \psi_B(t) = \delta(t) .
\end{equation}
The frequency filtering, however, has the effect that the photons
arriving at the APD has been delayed relative to their correlated twins
in the signal path. This can be taken into account by convoluting the
mode function with the filter response, which gives (assuming the
single Lorentzian filter)
\begin{equation}
  f_B(t) = \sqrt{2\kappa\eta_B} e^{\kappa t} \ ,\quad
    \text{for}\ t\le0\ \text{and 0 otherwise} .
\end{equation}
The overall trigger efficiency was also included in the filter
function.

Both filter functions assumes a photon click at time $t=0$. Due to the
inclusion of losses, they are not normalized to 1.

\section{IDEAL SQUEEZED QUBIT}

When we compare the experimentally obtained or theoretically predicted
states with the ideal squeezed qubit, we calculate the overlap
(fidelity) between the Wigner function of these states and the Wigner
function of the ideal squeezed qubit \cite{app:Note2}. Given parameters
$r$ (squeezing parameter), $\theta$ and $\phi$ (Bloch vector), the
ideal squeezed qubit has the state ket
\begin{equation}
  \ket{\mr{SQ}} =
    \cos\frac{\theta}{2} \hat S(r) \ket{0} + e^{i\varphi}
    \sin\frac{\theta}{2} \hat S(r) \ket{1} ,
\end{equation}
For the calculation of the Wigner function, $W\rs{SQ}(x,p) =
(2\pi)^{-1} \int e^{iyp} \langle x-y/2 \ket{\mr{SQ}} \bra{\mr{SQ}}
x+y/2 \rangle dy$, we need the wave functions of the squeezed vacuum
and the squeezed photon:
\begin{align}
  \bra{x} \hat S(r) \ket{0} &= (\pi e^{2r})^{-1/4} e^{-x^2/2e^{2r}} , \\
  \bra{x} \hat S(r) \ket{1} &=
      \bra{x} \frac{1}{\sinh r} \hat a \hat S(r) \ket{0} \nonumber \\
    &= \frac{1}{\sqrt{2}\sinh r} \left(x+\frac{d}{dx}\right)
      \bra{x} \hat S(r) \ket{0} \nonumber \\
    &= \sqrt{\frac{2}{\sqrt{\pi}e^{3r}}} \,x e^{-x^2/2e^{2r}} .
\end{align}

This gives the following expression for the squeezed qubit Wigner
function:
\begin{widetext}
\begin{equation}
  W\rs{SQ}(x,p) = \frac{1}{\pi} e^{-e^{-2r}x^2-e^{2r}p^2}
    \bigg(\! \cos\theta + (1-\cos\theta) \Big(\frac{x^2}{e^{2r}}
      + \frac{p^2}{e^{-2r}}\Big)
  + \sqrt{2}\Big(\cos\phi \frac{x}{e^r}+
    \sin\phi\frac{p}{e^{-r}}\Big)\sin\theta \bigg) .
\end{equation}
\end{widetext}

\end{document}